\documentstyle[titlepage,12pt]{article}

\def\be{\begin{equation}}
\def\bea{\begin{eqnarray}}

\def\grad{\nabla}
\def\ee{\end{equation}}
\def\eea{\end{eqnarray}}

\newcount\sectionnumber
\def\sect
{\global\equationnumber=0
\global\advance\sectionnumber by 1
\the\sectionnumber . }
\newcount\equationnumber
\def   \num
{\eqno{\global\advance\equationnumber by 1
\left(\the\sectionnumber .\the\equationnumber \right)}}

\begin{document}
\title{Quantum Evaporation of Liouville Black Holes}
\author{R.B. Mann \\
Department of Physics \\
University of Waterloo \\
Waterloo, Ontario \\
N2L 3G1}

\date{June 22, 1993\\
WATPHYS-TH93/01 \\
hep-th/9307072}

\begin{abstract}

The classical field equations of a Liouville field coupled to gravity in
two spacetime dimensions are shown to have black hole solutions.
Exact solutions are also obtained when quantum corrections due to back
reaction effects are included, modifying both the ADM mass and the black
hole entropy. The thermodynamic limit breaks down before evaporation of the
black hole is complete, indicating that higher-loop effects must be
included for a full description of the process. A scenario for the final
state of the black hole spacetime is suggested.
\end{abstract}

\maketitle

Liouville field theory  has been a useful tool in expanding our
understanding of 2D quantum gravity. The usual approach is to consider the
Liouville field to be the conformal factor of the metric, whose quantum
properties are then derived from the quantum dynamics of the Liouville
field coupled to other 2D matter \cite{Polya,DDK}.  Recently, however, a
different approach has been adopted in which  the Liouville field is taken
be an independent matter field whose stress-energy couples to 2D gravity in
a manner somewhat analogous to the $(3+1)$-dimensional case \cite{RtDil}.
The field equations contain asymptotically flat black hole solutions whose
temperature depends upon their ADM-mass, making them interesting objects of
investigation.

This letter presents the results of a study of the quantum dynamics of a
Liouville field in curved two-dimensional spacetime; details will appear in
a forthcoming paper \cite{Loubla}. Specifically, the classical field equations
of a
Liouville field in curved spacetime are modified to include quantum
corrections due to both conformally coupled matter and to the
gravity/Liouville system itself. An exact solution is obtained in each
case, thereby fully incorporating the back-reaction into the classical
black hole solution to one-loop order.

The action is a particular version of dilaton gravity \cite{MST,semi}
in which
\be
S_G + S_M = \frac{1}{2\pi}\int d^2x\sqrt{-g}\left(
\frac{1}{2} g^{\mu\nu}\grad_\mu\psi
   \grad_\nu\psi +\psi R \right) +S_M\label{1}
\ee
where $R$ is the Ricci scalar and $ S_M = \int d^2x\sqrt{-g}{\cal L}_M$ is
the two-dimensional matter action which is independent of the auxiliary
dilaton field $\psi$. This yields (after some manipulation) the field
equations
\be
R = 2\pi T_\mu^{\ \mu}  \qquad \qquad \grad_\nu T^{\mu\nu} = 0 \label{2}
\ee
\begin{equation}
\frac{1}{2}\left(\grad_\mu\psi\grad_\nu\psi - \frac{1}{2} g_{\mu\nu}
(\grad\psi)^2\right) + \frac{1}{2}
g_{\mu\nu}\grad^2\psi - \grad_\mu\grad_\nu\psi
= 2\pi (T_{\mu\nu}-\frac{1}{2}g_{\mu\nu}T^\lambda_{\lambda}) \label{3}
\end{equation}
where $T_{\mu\nu}$ is the conserved 2D stress-energy tensor associated with
$S_M$. Note that the evolution of the gravity/matter system is independent
of $\psi$.

Here the stress-energy is taken to be that of a Liouville field $\phi$,
so that $S_M = S_L$ where
\be
S_L = \frac{1}{2\pi}\int d^2x \sqrt{-g}\left( -\frac{1}{2} (\grad\phi)^2
+ \Lambda e^{-\phi} \right) \label{1a}
\ee
which reduces to that of the standard Liouville action for 2D flat space.
For simplicity a term coupling $\phi$ to the Ricci scalar has been set to
zero. In coordinates where $\partial/\partial t$ is a Killing vector
the classical field equations (\ref{2},\ref{3}) in this case
have the exact solution \cite{RtDil}
\be
ds^2 = -\alpha(x) dt^2 + \frac{dx^2}{\alpha(x)} \label{1b}
\ee
where
\begin{eqnarray}
\alpha(x) &=& (1-\frac{\Lambda}{4M^2}e^{-2Mx})
\nonumber\\
\phi = 2Mx &\mbox{and}& \psi = 2Mx + \psi_0 \label{4}
\end{eqnarray}
where $\psi_0$ and $M$ are constants of integration.
If $\Lambda > 0$ then the solution (\ref{4}) is that of an
asymptotically  flat black hole of ADM-mass $M$.

Although the black-hole metric in (\ref{4}) superficially resembles that
obtained in string-theoretic dilaton gravity, the mass parameter $M$ plays
a strikingly different role. Standard techniques for calculating the
temperature $T$ in the absence of back-reaction effects yield $T =
\frac{M}{2\pi}$ (and not $T=$constant \cite{CGHS}).
The black hole has positive specific heat and its entropy varies
logarithmically with the mass, a phenomenon found for other matter
couplings in the action (\ref{1}).

Consider next incorporating the back-reaction due to radiation of
conformally coupled matter fields with action $S_C$.
Classically, the addition of both (\ref{3}) and $S_C$ to
(\ref{1}) has no effect on the metric and Liouville fields as its
stress-energy tensor is traceless (although the dilaton field $\psi$ is
modified).  Quantum-mechanically, however, the trace anomaly
$\langle O\vert T_{\mu}^{c\ \mu}\vert O\rangle
= -\frac{c_M}{24\pi}R$,
where $c_M$ is the central charge, implies \cite{TomRobb}
\begin{eqnarray}
\langle O|T^c_{11}| O\rangle = \frac{1}{96\pi}
\frac{(\alpha^\prime)^2 - T_0}{\alpha^2} & \quad &
\langle O|T^c_{01}| O\rangle = \frac{T_1}{\alpha^2}
\nonumber \\
\langle O|T^c_{00}|O\rangle &=& \frac{1}{96\pi}
\frac{4\alpha^{\prime\prime}-(\alpha^\prime)^2 + T_0}{\alpha^2}
\qquad\qquad\label{6}
\end{eqnarray}
where $\alpha^\prime \equiv \frac{d}{dx}\alpha$. $T_0$ and $T_1$ are
constants which depend upon the choice of boundary conditions and
$T^c_{\mu\nu}$ is the stress-energy tensor of the conformally coupled
matter.

The field equations (\ref{2},\ref{3}) of the dilaton/metric/Liouville
system are consequently modified. These have the exact solution
\begin{eqnarray}
\alpha_B &=& 2\hat{M}x - \frac{\Lambda}{2}\frac{c_M^2}{(24-c_M)(12-c_M)}
x_0^2(\frac{x}{x_0})^{-24/c_M} \nonumber\\
\phi_B &=& (1-\frac{12}{c_M})\ln(\frac{x}{x_0}) \quad\mbox{and}\quad
\psi_B = -\frac{24}{c_M}\ln(\frac{x}{x_0}) +\psi_0 \label{5}
\end{eqnarray}
where  $T_1=0$ and $T_0 = 4M^2$,  and the subscript ``B'' denotes the fact
that the  back-reaction has been included.  Here $\hat{M} =
M\frac{c_M}{c_M-24}$, and $\psi_0$ and $x_0$ are constants  of integration,
the latter chosen so that the Liouville field vanishes at the horizon. The
ADM-mass is now modified
\be
M_B = M \frac{(1-\frac{c_M}{12})^2}{1-\frac{c_M}{24}}
\label{6a}
\ee
due to the presence of the quantum stress-energy tensor.

There is a singularity in the model for $c_M=12$, where the field equations
are ill-defined. For $c_M<0$ or $24 > c_M > 12$ the solution (\ref{5}) is
that of an asymptotically flat ($\lim_{x\to\infty}R = 0$)
black hole of positive mass $M_B$ as given by (\ref{6}), with
temperature $T=\frac{M}{2\pi}$ and entropy
\be
S = \frac{(1-\frac{c_M}{12})^2}{(1-\frac{c_M}{24})}
\ln\left(\frac{M_B}{M_0}\right)
\label{7}
\ee
where $M_0$ is a constant of integration. For increasingly large negative
$c_M$ (or for $24 > c_M > 12$) the density of states increases
exponentially with increasing $|c_M|$.

For $12 > c_M > 0$, the metric in (\ref{5}) becomes that of an expanding
2D universe with time coordinate $x=\tilde{t}$. The Ricci scalar $R\sim
\tilde{t}^{\frac{24}{c_M}-2}$ and is finite for small $\tilde{t}$. For
$c_M > 24$ the metric is still that of an asymptotically flat black hole,
but with negative ADM-mass.

The situation changes somewhat when quantum corrections due to the
dilaton/metric/Liouville system itself are included. This may be done
by  considering a functional integral over the field configurations of the
metric, $\psi$, and $\phi$ fields \cite{2dqgm}.
The path integral is
\be
Z=\int \frac{{\cal D}g}{V_{GC}}
{\cal D}\psi {\cal D}\phi {\cal D}\Phi e^{-(S[\psi,g,\phi] +S_M[\Phi])\hbar}
\label{25q}
\ee
where $S[\psi,g,\phi] = S_G +S_L$  is the Euclideanization of
(\ref{1},\ref{1a}) and $S_M$ the part of the action incorporating
additional matter fields $\{\Phi\}$. The volume of the diffeomorphism
group, $V_{GC}$, has been factored out.

Making the same scaling assumption as in refs. \cite{DDK,2dqgm} about
the functional measure  yields
\begin{eqnarray}
Z&=&\int [{\cal D}\tau] {\cal D}_g\phi {{\cal D}_g b} {{\cal D}_g c}
{\cal D}_g\psi {\cal D}_g\phi{\cal D}_g\Phi
e^{-(S[\psi,\rho,\phi]+S_{gh}[b,c]+S_M[\Phi])/\hbar}
\label{26q}\\
&=&\int [{\cal D}\tau] {\cal D}_{\hat{g}}\phi {{\cal D}_{\hat{g}} b}
{{\cal D}_{\hat{g}} c} {\cal D}_{\hat{g}}\psi {\cal D}_{\hat{g}}\phi
{\cal D}_{\hat{g}}\Phi
e^{-(S[\psi,\rho,\phi]+S_{gh}[b,c]+S_M[\Phi]+\hat{S}[\rho,\hat{g}])/\hbar}
\nonumber
\end{eqnarray}
where $[{\cal D}\tau]$ represents the integration over the Teichmuller
parameters and
\be
\hat{S}[\rho,\hat{g}]=\frac{1}{8\pi}\int d^2x \sqrt{\hat{g}}\left(
\hat{g}^{\mu\nu}\partial_\mu\rho\partial_\nu\rho - Q \rho\hat{R} \right)
\label{27q}
\ee
is the Liouville action with arbitrary coefficient $Q$, where $g_{\mu\nu}=
e^{\beta\rho}\hat{g}_{\mu\nu}$, and $\hat{R}$  and
$\hat{\grad}^2$  are respectively the curvature scalar and Laplacian of the
metric  $\hat{g}$. $S_{gh}$ is the action for the ghost fields $b$ and $c$.

Determining $\beta$ and $Q$ from the requirement that the conformal anomaly
vanish and that the gravitationally dressed Liouville potential
$\Lambda e^{\beta\rho-\phi}$ is a conformal tensor of weight (1,1)
yields
\be
\beta^2 = \frac{47-c_M}{2c_M-91} \label{11}
\ee
and $Q = 3\beta$,
leading to the restriction $47 > c_M > 91/2$,
where $c_M$ is the central charge of the conformally coupled
matter fields. The cosmological constant term has been renormalized to
zero.

In this case, the
modified field equations have the exact solution
\begin{eqnarray}
\alpha_Q &=& 2\tilde{M}x -
\frac{\Lambda}{2}\frac{\beta^2+1}{(2\beta^2+1)}
x_0^2(\frac{x}{x_0})^{-2\beta^2} \nonumber\\
\phi_Q &=& (1+\beta^2)\ln(\frac{x}{x_0})
\quad\mbox{and}\quad  \psi_B = -2\beta^2\ln(\frac{x}{x_0}) +\psi_0 \label{5a}
\end{eqnarray}
where $\tilde{M} = \frac{M}{1+2\beta^2}$ and
the subscript ``Q'' denotes the fact that the fully quantized
back-reaction has been included.  The ADM-mass is now
\be
M_Q = 2M \frac{(1+\beta^2)^2}{\beta^2(1+2\beta^2)}
\label{6b}
\ee
and is manifestly positive.

The spacetime described by the metric in (\ref{5a}) is that of a positive
mass black hole. It is asymptotically flat
(more properly, asymptotically Rindlerian) for large positive $x$, and has
a singularity in the curvature, dilaton and Liouville fields at $x=0$.
The entropy is
\be
S = 2\frac{(\beta^2+1)^2}{\beta^2(2\beta^2+1)}
\ln\left(\frac{M_Q}{M_0}\right)
\label{7z}
\ee
the temperature again being given by $T=\frac{M}{2\pi}$.

The above considerations suggest the following scenario for the evaporation
of Liouville black holes. When quantum effects are taken into account, the
classical spacetime described by (\ref{4}) is modified to (\ref{5a}), with
a larger ADM-mass due to inclusion of the quantum stress-energy.  This mass
will decrease with time as $M_Q \sim 1/t$, decreasing both the location
$x_H$ of the horizon and the maximal proper distance from the horizon
to the singularity. Eventually this distance (the ``size'' of the
event horizon) becomes comparable to the compton wavelength, $x_H \sim
1/M_Q$, or
\be
M_Q \ge \sqrt{\lambda\frac{1+\beta^2}{\beta^2(1+2\beta^2)}}
\equiv M_0
\label{8}
\ee
which may be taken as a definition of $M_0$.   Since  as $M_Q \to M_0$ the
entropy tends to zero for finite temperature, at this point the
thermodynamic description breaks down and higher-loop corrections become
important.

What effect might these have? Previous work \cite{MST} suggests that
quantum vacuum energies  $\Lambda_Q$ (renormalized to zero to this order)
will modify the temperature  to ${T} \sim \sqrt{M^2 -
\frac{\Lambda_Q}{2}}$, so that entropy becomes
$$
S \sim \ln\left( \frac{\sqrt{M^2 - \frac{\Lambda_Q}{2}} +
M}{\tilde{M}_0}\right)
$$
where $\tilde{M}_0$ is a constant comparable in magnitude to $M_0$. In this
case the black hole slowly cools off to a zero temperature remnant, leaving
behind a global event horizon with its requisite loss of quantum coherence.
Work is in progress to determine whether or not  higher-loop effects will
have similar consequences.

\section*{Acknowledgements}
I am grateful to Dr. M. Morris for discussions.
This work was supported by the Natural Sciences and Engineering Research
Council of Canada.


\begin{thebibliography}{References}

\bibitem{Polya}A.M. Polyakov, Phys. Lett {\bf B103} (1981) 207.
\bibitem{DDK}F. David, Mod. Phys. Lett. {\bf A3} (1988) 1651; J. Distler
and H. Kawai, Nucl. Phys. {\bf B321} (1989) 509.
\bibitem{RtDil}R.B. Mann, Phys. Rev. {\bf D47} (1993) 4438.
\bibitem{Loubla}R.B. Mann, WATPHYS TH-93/02, in preparation.
\bibitem{MST}R.B. Mann, A. Shiekh, and L. Tarasov, Nucl. Phys. {\bf B341}
(1990) 134.
\bibitem{semi}R.B. Mann, S.M. Morsink, A.E. Sikkema and T.G. Steele,
Phys. Rev. {\bf D43} (1991) 3948.
\bibitem{CGHS}C. Callan, S. Giddings, J. Harvey and A. Strominger,
Phys. Rev. {\bf D45} (1992) R1005 .
\bibitem{TomRobb}R.B. Mann and T.G. Steele, Class. Quant. Grav. {\bf 9}
(1992) 475.
\bibitem{2dqgm}R.B. Mann, Phys. Lett. {\bf B294} (1992) 310.

 \end{thebibliography}
 \end{document}